\documentclass[copyright,creativecommons]{eptcs}
\providecommand{\event}{HOR 2010} % Name of the event you are submitting to

\usepackage{breakurl}             % Not needed if you use pdflatex only.
\usepackage[spanish, english, activeacute]{babel}
\usepackage{amsmath}
\usepackage{amsfonts}
\usepackage{amssymb}
\usepackage{amsthm}
\usepackage{tikz}
\usepackage{colonequals}
\usepackage{fancybox}
\usepackage{hyperref}
\usepackage{subfigure}
\usepackage{wrapfig}
\usepackage{enumerate}

\usetikzlibrary{matrix}
\tikzstyle flechas locas=[color=gray!90,thick]

\newcommand{\calculusbox}{\fbox}

\newcommand{\stepjust}[2]{\underset{\mbox{\tiny{#1}}}{#2}}
\newcommand{\ie}[1]{\emph{i.e.,} #1}

\newcommand{\esabbr}{ES}

\newcommand{\oursection}[1]{
	\vspace{-.3cm}
	\section{#1}
	\vspace{-.2cm}
}

\newcommand{\oursectionshort}[2]{
	\vspace{-.3cm}
	\section[#2]{#1}
	\vspace{-.2cm}
}

\newcommand{\oursubsection}[1]{
	\vspace{-.1cm}
	\subsection{#1}
	\vspace{-.1cm}
}

\theoremstyle{definition}
\newtheorem{definition}{Definition}%[chapter]
\newtheorem{theorem}[definition]{Theorem}
\newtheorem*{theorem*}{Theorem}
\newtheorem{lemma}[definition]{Lemma}
\newtheorem{cor}[definition]{Corollary}

\newtheorem*{obs*}{Observation}

\theoremstyle{plain}
\newtheorem*{note*}{Note}

\newcommand{\naturals}{\mathbb{N}}
\newcommand{\naturalsnonzero}{\mathbb{N}_{>0}}

\newcommand{\powerset}[1]{\mathcal{P}\!\left( #1 \right)}

\newcommand{\en}{\rightarrow}

\newcommand{\varset}[3]{\left\{ #1_{#2} , \ldots , #1_{#3} \right\}}
\newcommand{\numset}[2]{\left\{ #1 , \ldots , #2 \right\}}

\newcommand{\vlste}[3]{\left[ {#1}_{#2}, \ldots, {#1}_{#3} \right]}

\newcommand{\comp}[2]{#1 \circ #2}

\newcommand{\funcph}{\bullet}

\newcommand{\setl}[2]{#1_{< #2}}
\newcommand{\setleq}[2]{#1_{\leq #2}}
\newcommand{\setg}[2]{#1_{> #2}}

\newcommand{\seteq}[2]{#1_{= #2}}

\newcommand{\setoplus}[2]{#1_{\oplus \, #2}}

\newcommand{\fvLone}{\textrm{FV}}
\newcommand{\fv}[1]{\fvLone \! \left( #1 \right)}

\newcommand{\abs}[1]{\lambda #1}
\newcommand{\absv}[2]{\lambda #1 . #2}
\newcommand{\app}[2]{#1\,#2}

\newcommand{\ctxt}[2]{ #1\left[ #2 \right] }

\newcommand{\bnfeq}{\coloncolonequals}
\newcommand{\alphaequiv}{=_{\alpha} \!}
\newcommand{\orspaced}{\ |\ }
\newcommand{\reducea}{\rightarrow}

\newcommand{\mvar}[2]{\textrm{#1}_{#2}}

\newcommand{\lambdaDB}{\lambda_{\textrm{dB}}}
\newcommand{\lambdaDBterm}{\Lambda_\textrm{dB}}

\newcommand{\lambdaT}{\lambda \textrm{t}}

\newcommand{\lambdaR}{\lambda{\textrm{r}}}

\newcommand{\RE}{\textrm{re}}
\newcommand{\lambdaRE}{\lambda{\RE}}
\newcommand{\lambdaREterm}{\Lambda{\RE}}
\newcommand{\lambdaREOPterm}{\Lambda{\RE}_{\textrm{op}}}

\newcommand{\REGC}{\textrm{re}_{\textrm{gc}}}
\newcommand{\lambdaREGC}{\lambda{\REGC}}

\newcommand{\REXP}{\textrm{rex}_\textrm{p}}
\newcommand{\lambdaREXP}{\lambda{\REXP}}
\newcommand{\REX}{\textrm{rex}}
\newcommand{\lambdaREX}{\lambda{\REX}}

\newcommand{\lambdaX}{\lambda{\textrm{x}}}
\newcommand{\lambdaXterm}{\Lambda{\textrm{x}}}
\newcommand{\lambdaXOPterm}{\Lambda{\textrm{x}}_{\textrm{op}}}

\newcommand{\lambdaS}{\lambda \textrm{s}}

\newcommand{\lambdaSE}{\lambda \textrm{s}_e}

\newcommand{\EX}{\textrm{ex}}
\newcommand{\lambdaEX}{\lambda{\EX}}
\newcommand{\BX}{\textrm{Bx}}

\newcommand{\XGC}{\textrm{xgc}}
\newcommand{\lambdaXGC}{\lambda{\XGC}}

\newcommand{\lambdaSigma}{\lambda\sigma}
\newcommand{\lambdaSigmaLift}{\lambda\sigma_\Uparrow}

\newcommand{\lambdaCXiPhi}{C \lambda \xi \phi}
\newcommand{\lambdaXiPhi}{\lambda \xi \phi}
\newcommand{\lambdaUpsilon}{\lambda \upsilon}

\newcommand{\lift}[1]{\Uparrow \! ({#1})}
\newcommand{\shift}{\uparrow}
\newcommand{\conssigma}[2]{{#1} \, \cdot {#2}}
\newcommand{\compsigma}[2]{{#1} \, \circ {#2}}
\newcommand{\megaSwap}[1]{\Updownarrow \! ({#1})}

\newcommand{\tradWLone}[1]{\textrm{w}_{#1}}
\newcommand{\tradW}[2]{\textrm{w}_{#1} \! \left( #2 \right)}
\newcommand{\tradULone}[1]{\textrm{u}_{#1}}
\newcommand{\tradU}[2]{\textrm{u}_{#1} \! \left( #2 \right)}

\newcommand{\tradWUniLone}{\textrm{w}}
\newcommand{\tradWUni}[1]{\textrm{w} \! \left( #1 \right)}
\newcommand{\tradUUniLone}{\textrm{u}}
\newcommand{\tradUUni}[1]{\textrm{u} \! \left( #1 \right)}

\newcommand{\lambdasust}[3]{#1 \! \left\{ #2 \colonequals #3 \right\}}

\newcommand{\updLone}[2]{\mathrm{U}_{#1}^{#2}}
\newcommand{\upd}[3]{\mathrm{U}_{#1}^{#2}({#3})}
\newcommand{\lambdaDBsust}[3]{#1\{\!\!\{ #2 \leftarrow #3 \}\!\!\}}
\newcommand{\betaDB}{\beta_{\mathrm{dB}}}
\newcommand{\betaDBstep}{\ \reducea_{\betaDB}\ }

\newcommand{\lambdaXsust}[3]{#1 [ #2 \colonequals #3 ]}

\newcommand{\BXstep}{\reducea_{\BX}}

\newcommand{\cequiv}{=_{\mathrm{C}}}

\newcommand{\addOneLone}[1]{\uparrow_{#1}}
\newcommand{\addOne}[2]{\uparrow_{#1}\!\!({#2})}
\newcommand{\addOneChainLone}[1]{\Uparrow^{#1}}
\newcommand{\addOneChain}[2]{\Uparrow^{#1}\!\!({#2})}
\newcommand{\swapLone}[1]{\updownarrow_{#1}}
\newcommand{\swap}[2]{\updownarrow_{#1}\!\!({#2})}
\newcommand{\swapChainLone}[2]{\Updownarrow_{#1}^{#2}}
\newcommand{\swapChain}[3]{\Updownarrow_{#1}^{#2}\!\!({#3})}
\newcommand{\lambdaRsust}[2]{#1\{#2\}}
\newcommand{\betaR}{\beta_\textrm{r}}
\newcommand{\betaRstep}{\reducea_{\betaR}}

\newcommand{\LambdaREsustalt}{\sigma}
\newcommand{\LambdaREsust}[2]{\LambdaREsustalt^{#2} #1}
\newcommand{\lambdaREsust}[2]{#1 \! \left[ #2 \right]}

\newcommand{\subOneLone}[1]{\downarrow_{#1}}
\newcommand{\subOne}[2]{\downarrow_{#1}\!\!({#2})}

\newcommand{\lambdaREXPstep}{\reducea_{\lambdaREXP}}

\newcommand{\eqdPlain}{EqD}

\newcommand{\dequiv}{=_{\mathrm{D}}}

\newenvironment{trs}{\begin{center}\begin{tabular}{llcll}}{\end{tabular}\end{center}}

\newcommand{\trslrule}[3]{(#1) & $#2$ & $\reducea$ & $#3$ &\\}
\newcommand{\trslcrule}[4]{(#1) & $#2$ & $\reducea$ & $#3$ & (#4)\\}
\newcommand{\trslceq}[4]{(#1) & $#2$ & $=$ & $#3$ & (#4)\\}

\newcommand{\trsblank}{\\}

\title{Swapping: a natural bridge between named and indexed explicit substitution calculi}
\author{
Ariel Mendelzon \\
\institute{Depto. de Computaci\'on, FCEyN, \\
    Universidad de Buenos Aires.}
\email{amendelzon@dc.uba.ar}
\and
Alejandro R\'ios \\
\institute{Depto. de Computaci\'on, FCEyN, \\
    Universidad de Buenos Aires.}
\email{rios@dc.uba.ar}
\and
Beta Ziliani \\
\institute{Depto. de Computaci\'on, FCEyN, \\
    Universidad de Buenos Aires.}
\email{lziliani@dc.uba.ar}
}
\def\titlerunning{Swapping: a natural bridge between named and indexed explicit substitution calculi}
\def\authorrunning{Ariel Mendelzon, Alejandro R\'ios \& Beta Ziliani}
\begin{document}
\maketitle

% Abstract
\vspace{-.1cm}
\begin{abstract}
This article is devoted to the presentation of $\lambdaREX$, an explicit
substitution calculus with \emph{de Bruijn} indexes and a simple notation.
By being isomorphic to $\lambdaEX$ \mbox{-- a recent formalism with variable names --,}
$\lambdaREX$ accomplishes simulation of $\beta$-reduction (Sim), preservation of
$\beta$-strong normalization (PSN) and meta-confluence (MC), among
other desirable properties.
Our calculus is based on a novel presentation of $\lambdaDB$,
using a \emph{swap} notion that was originally devised by de Bruijn.
Besides $\lambdaREX$, two other indexed calculi isomorphic to
$\lambdaX$ and $\lambdaXGC$ are presented,
demonstrating the potential of our technique when applied to the
design of indexed versions of known named calculi.
\end{abstract}
\vspace{-.1cm}

% Introduction
\oursection{Introduction}

This article is devoted to explicit substitutions (\esabbr, for short), a
formalism that has attracted attention since the appearance of $\lambdaSigma$
\cite{AbCaCuLe1990} and, later, of Melli\`{e}s' counterexample \cite{Me1995}, showing
the lack of the preservation of $\beta$-strong normalization property (PSN, for short) 
in $\lambdaSigma$. One of the main motivations behind the field of \esabbr\ is 
studying how substitution behaves when internalized in the language it serves (in the classic $\lambda$-calculus,
substitution is a meta-level operation). Several calculi have been proposed
since the counterexample of Melli\`{e}s, and few have been shown to have a
whole set of desirable properties: simulation of $\beta$-reduction, PSN, meta-confluence, full composition, etc. For a
detailed introduction to the \esabbr\ field, we refer the reader to e.g.
\cite{Le1994,Ke2009,RoBlLa20009}.

In 2008, D. Kesner proposed $\lambdaEX$ \cite{Ke2008,Ke2009}, a formalism with
variable names that has the entire set of properties expected from an \esabbr\
calculus. As Kesner points in \cite{Ke2009}, for implementation purposes a different approach
to variable names should be taken, since bound variable renaming (\ie{working modulo $\alpha$-equivalence}) is
known to be error-prone and computationally expensive.
Among others, one of the ways this problem is tackled is by using \emph{de Bruijn} notation \cite{Br1972}, 
which is a technique that simply avoids the need of working modulo $\alpha$-equivalence.
As far as we know, no \esabbr\ calculus with \emph{de Bruijn} indexes
and the whole set of properties enjoyed by $\lambdaEX$ exists to date. 
The main target of this article is the introduction of $\lambdaREX$,
an \esabbr\ calculus with \emph{de Bruijn} indexes that, by being isomorphic to $\lambdaEX$, enjoys the
same set of properties. $\lambdaREX$ is based on $\lambdaR$, a novel
\emph{swapping}-based version of the classic $\lambdaDB$ \cite{Br1972}, that
we also introduce here.

It is important to remark that the whole development was made on a staged
basis: we first devised $\lambdaR$, and then made substitutions explicit
orienting the definition for $\lambdaR$'s meta-substitution. At that point, we
got a calculus we called $\lambdaRE$, which turned out to be isomorphic to
$\lambdaX$ \cite{BlRo1995,BlGe1997}. Encouraged by this result, we added
\emph{Garbage Collection} to $\lambdaRE$, obtaining a calculus isomorphic 
to $\lambdaXGC$ \cite{BlRo1995}: $\lambdaREGC$. 
Finally, we added composition of substitutions in the style
of $\lambdaEX$ to $\lambdaREGC$, obtaining $\lambdaREX$. 
Thus, besides fulfilling our original aim, we introduced
\emph{swapping}, a technique that turns out
to behave as a natural bridge between named and indexed
formalisms. Furthermore, we didn't know any indexed 
isomorphic versions of $\lambdaX$ nor $\lambdaXGC$.

The content of the article is as follows: in Section \ref{sec:lambda_r} we
present $\lambdaR$, an alternative version of $\lambdaDB$. Next, in Section
\ref{sec:lambda_rex}, we introduce $\lambdaRE$, $\lambdaREGC$ and $\lambdaREX$,
the three \esabbr\ calculi derived from $\lambdaR$ already mentioned. We show
the isomorphism between these and the $\lambdaX$, $\lambdaXGC$ and $\lambdaEX$
calculi in Section \ref{sec:isom}. Last, in sections \ref{sec:related_work} and
\ref{sec:conclusions}, we point out related work and present the conclusions,
respectively.  We refer the interested reader to \cite{Me2010} for complete
proofs over the whole development.

% Lambda_r part
\oursection{A new presentation for $\lambdaDB$: the $\lambdaR$-calculus}\label{sec:lambda_r}

\oursubsection{Intuition}

The $\lambda$-calculus with \emph{de Bruijn} indexes ($\lambdaDB$, for short)
\cite{Br1972} accomplishes the elimination of $\alpha$-equivalence, since
$\alpha$-equivalent $\lambda$-terms are syntactically identical under
$\lambdaDB$. This greatly simplifies implementations, since caring about bound
variable renaming is no longer necessary. One usually refers to a \emph{de
Bruijn} indexed calculus as a \emph{nameless calculus}, for binding is
positional -- relative -- instead of absolute (indexes are used in place of
names for this purpose).  We observe here that, even though this
\emph{nameless} notion makes sense in the classical $\lambdaDB$-calculus
(because the substitution operator is located in the meta-level), it seems not
to be the case in certain \esabbr\ calculi derived from $\lambdaDB$, such as:
$\lambdaS$ \cite{KaRi1995}, $\lambdaSE$ \cite{KaRi1997} or $\lambdaT$
\cite{KaRi1998}.  These calculi have constructions of the form
$\lambdaXsust{a}{i}{b}$ to denote \esabbr\ (notations vary). Here, even though
$i$ is not a name \emph{per se}, it plays a similar role: $i$ indicates which
free variable should be substituted; then, these calculi are not purely
\emph{nameless}, \ie{binding is \emph{mixed}: positional (relative) for
abstractions and named (absolute) for closures}.

In general, we observe that not a single \esabbr\ calculus with \emph{de
Bruijn} indexes to date is completely nameless.  This assertion rests on the
following observation: in each and every case, the (Lamb) rule is of the form
$\lambdaREsust{(\abs{a})}{s} \reducea \abs{\lambdaREsust{a}{s'}}$. Thus, since
the term $a$ is not altered, an ``absolute binding technique'' \emph{must} be
implemented inside $s$ in order to indicate which free variable is to be
substituted. To further support this \emph{not-completely-nameless} assertion,
we note that even though there is a known isomorphism between the classic
$\lambda$-calculus and the $\lambdaDB$-calculus, when substitutions are made
explicit in both calculi, the isomorphism does not hold just by adding the new
\esabbr\ case (which would be reasonable to expect). The problem is that
$\lambdaDB$'s classic definition is always -- tacitly, at least -- being used
for the explicitation task, thus obtaining calculi with mixed binding
approaches, as mentioned earlier. As shown throughout the rest of the paper,
our (Lamb) rules will be of the form $\lambdaREsust{(\abs{a})}{s} \reducea
\abs{\lambdaREsust{a'}{s'}}$, \ie{altering \emph{both} $a$ \emph{and} $s$} to
enforce a \emph{completely nameless} approach.

\begin{wrapfigure}{r}{0.22\textwidth}
\centering
\vspace{-10pt}
\subfigure[]{\label{lambda_r:fig:bindings:before}
    \begin{tikzpicture}
        \draw
            (0,0) node(sust) {$\LambdaREsustalt^b$}
            (0.25,0) node {$($}
            (0.5,0) node(abs) {$\lambda$}
            (1,0) node(v1)  {$1$}
            (1.5,0) node(v2)  {$2$}
            (1.75,0) node {$)$};
        \draw[<-,style=flechas locas] (sust.south) to [out=-25,in=-155] (v2.south);
        \draw[<-,style=flechas locas] (abs.north) to [out=25,in=155] (v1.north);
    \end{tikzpicture}
}
\vfill
\subfigure[]{\label{lambda_r:fig:bindings:after}
    \begin{tikzpicture}
        \draw
            (0,0) node(abs) {$\lambda$}
            (0.25,0) node {$($}
            (0.5,0) node(sust) {$\LambdaREsustalt^b$}
            (1,0) node(v1)  {$1$}
            (1.5,0) node(v2)  {$2$}
            (1.75,0) node {$)$};
        \draw[<-,style=flechas locas] (abs.south) to [out=-25,in=-155] (v2.south);
        \draw[<-,style=flechas locas] (sust.north) to [out=25,in=155] (v1.north);
    \end{tikzpicture}
}
\vfill
\subfigure[]{\label{lambda_r:fig:bindings:after_swap}
    \begin{tikzpicture}
        \draw
            (0,0) node(abs) {$\lambda$}
            (0.25,0) node {$($}
            (0.5,0) node(sust) {$\LambdaREsustalt^b$}
            (1,0) node(v2)  {$2$}
            (1.5,0) node(v1)  {$1$}
            (1.75,0) node {$)$};
        \draw[<-,style=flechas locas] (abs.south) to [out=-25,in=-155] (v2.south);
        \draw[<-,style=flechas locas] (sust.north) to [out=25,in=155] (v1.north);
    \end{tikzpicture}
}
\caption{Bindings}
\label{lambda_r:fig:bindings}
\end{wrapfigure}
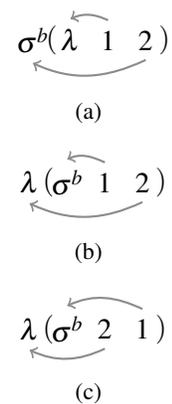

In order to obtain a \emph{completely nameless} notion for an explicit substitutions
$\lambdaDB$, we start by eliminating the index $i$ from the substitution
operator. Then, we are left with terms of the form $\lambdaREsust{a}{b}$, and
with a (Beta) reduction rule that changes from $\; \app{(\abs{a})}{b} \reducea
\lambdaXsust{a}{1}{b} \;$ to $\; \app{(\abs{a})}{b} \reducea
\lambdaREsust{a}{b}$. The semantics of $\lambdaREsust{a}{b}$ should be clear
from the new (Beta) rule. The problem is, of course, how to define it. Two
difficulties arise when a substitution crosses (goes into) an abstraction:
first, the indexes of $b$ should be incremented in order to reflect the new
variable bindings; second -- and the key to our technology --, some mechanism
should be implemented in order to replace the need for indexes inside closures
(since these should be incremented, too).

The first problem is solved easily: we just use an operator to progressively
increment indexes with every abstraction crossing, in the style of $\lambdaT$
\cite{KaRi1998}. The second issue is a bit harder.  Figure
\ref{lambda_r:fig:bindings} will help us clarify what we do when a substitution
crosses an abstraction, momentarily using $\; \LambdaREsust{a}{b} \;$ to denote
$\; \lambdaREsust{a}{b} \;$ in order to emphasize the binding character of the
substitution (by writing the substitution construction \emph{before} the term
and \emph{annotating} it with the substituent -- which does not actually affect
binding --, it resembles the abstraction operation; thus, ``reading'' the term
is much easier for those who are already familiar with \emph{de Bruijn}
notation). In this example we use the term $\;
\LambdaREsust{(\abs{\app{1}{2}})}{b}$ (which stands for $\;
\lambdaREsust{(\abs{\app{1}{2}})}{b}$). Figure
\ref{lambda_r:fig:bindings:before} shows the bindings in the original term;
Figure \ref{lambda_r:fig:bindings:after} shows that bindings are inverted if we
cross the abstraction and do not make any changes. Then, in order to get
bindings ``back on the road'', we just \emph{swap} indexes $1$ and $2$! (Figure
\ref{lambda_r:fig:bindings:after_swap}). With this operation we recover,
intuitively, the original semantics of the term. Summarizing, all that is
needed when abstractions are crossed is: \emph{swap} indexes $1$ and $2$ and,
also, increment the indexes of the term carried in the substitution. That is
exactly what $\lambdaR$ does, with substitutions in the meta-level. 

In Section \ref{lambda_r:sec:def} we define both $\lambdaDB$ and $\lambdaR$; in
Section \ref{lambda_r:sec:igualdad_lambda_db} we show that they are the same
calculus.

%Next,
%we give definitions for both $\lambdaDB$ and $\lambdaR$. , in order to show their
%equivalence.

\oursubsection{Definitions}\label{lambda_r:sec:def}

First of all, we define some operations on sets of naturals numbers.

\begin{definition}[Operations on sets of natural numbers]
For every $N \subset \naturals$, $k \in \naturals$:
\begin{enumerate}
    \item $N + \, k            		= \left\{ n + k : n \in N \right\}$
    \item $N - \, k        				= \left\{ n - k : n \in N \land n > k \right\}$
    \item $\setoplus{N}{k}     		= \left\{ n : n \in N \land n \oplus k \right\}$, with $\oplus \in \left\{ =, <, \leq, >, \geq \right\}$
\end{enumerate}
\end{definition}

\vspace{.3cm}
Terms for $\lambdaR$ are the same as those for $\lambdaDB$. That is:

\begin{definition}[Terms for $\lambdaDB$ and $\lambdaR$]
The set of terms for $\lambdaDB$ and $\lambdaR$, denoted $\lambdaDBterm$, is given in BNF by:
\vspace{-.15cm}
\[
    a \bnfeq n \orspaced \app{a}{a} \orspaced \abs{a} \qquad (n \in \naturalsnonzero)
    \vspace{-.15cm}
\]
\end{definition}

\begin{definition}[Free variables]
The free variables of a term, $\fvLone :
\lambdaDBterm \en \powerset{\naturalsnonzero}$, is given by:
\[
    \fv{n}                      =   \left\{ n \right\} \qquad
    \fv{\app{a}{b}}             =   \fv{a} \cup \fv{b} \qquad
    \fv{\abs{a}}                =   \fv{a} - 1
\]
\end{definition}

\subsubsection*{Classical definitions}

We recall the classical definitions for $\lambdaDB$ (see e.g. \cite{KaRi1995}
for a more detailed introduction).

\begin{definition}[Updating meta-operator for $\lambdaDB$]
For every $k \in \naturals$, $i \in \naturalsnonzero$,
$\updLone{k}{i} : \lambdaDBterm \en \lambdaDBterm$ is given inductively by:
\[
    \begin{array}{lll}
        \upd{k}{i}{n} & = &   
            \left\{ 
            \begin{array}{ll}
                n               &   \mbox{if } n \leq k \\
                n + i - 1       &   \mbox{if } n > k
            \end{array} 
            \right. \\
    \end{array}
    \qquad
    \begin{array}{lll}
        \upd{k}{i}{\app{a}{b}}    & = &   \app{\upd{k}{i}{a}}{\upd{k}{i}{b}} \\
        \upd{k}{i}{\abs{a}}       & = &   \abs{\upd{k+1}{i}{a}} 
    \end{array}
\]
\end{definition}

\begin{definition}[Meta-substitution for $\lambdaDB$]
For every $a, b, c \in \lambdaDBterm$, $m, n \in \naturalsnonzero$,
$\lambdaDBsust{\funcph}{\funcph}{\funcph} : \lambdaDBterm \times \naturalsnonzero \times \lambdaDBterm \en
\lambdaDBterm$ is given inductively by:
\[
    \begin{array}{lll}
        \lambdaDBsust{m}{n}{c}  & = & 
            \left\{ 
            \begin{array}{ll}
                m                   &   \mbox{if } m < n \\
                \upd{0}{n}{c}       &   \mbox{if } m = n \\
                m-1                 &   \mbox{if } m > n \\
            \end{array} 
            \right. \\
    \end{array}
    \qquad
    \begin{array}{lll}
        \lambdaDBsust{(\app{a}{b})}{n}{c}           & = &   \app{\lambdaDBsust{a}{n}{c}}{\lambdaDBsust{b}{n}{c}} \\
        \lambdaDBsust{(\abs{a})}{n}{c}                 & = &   \abs{\lambdaDBsust{a}{n+1}{c}}
    \end{array}
\]
\end{definition}

\begin{definition}[$\lambdaDB$-calculus] 
The $\lambdaDB$-calculus is the reduction system $\left( \lambdaDBterm, \betaDB
\right)$, where:
\[
    \left( \forall a,b \in \lambdaDBterm \right) \left( a \betaDBstep b\ \iff \left( \exists \ C\ \mbox{context};\ c, d \in \lambdaDBterm \right) 
    \left( a = \ctxt{\mathrm{C}}{(\abs{c})d} \land b = \ctxt{\mathrm{C}}{\lambdaDBsust{c}{1}{d}} \right) \right)
\]
\end{definition}

\pagebreak[3]

\vspace{-.5cm}
\subsubsection*{New definitions}

We now define the new meta-operators used to implement index
increments and swaps.

\begin{definition}[\emph{Increment operator} -- $\addOneLone{i}$] 
For every $i \in \naturals$, $\addOneLone{i}\ : \lambdaDBterm \en
\lambdaDBterm$ is given inductively by:
\[
    \begin{array}{lll}
        \addOne{i}{n} & = &   
            \left\{ 
            \begin{array}{ll}
                n               &   \mbox{if} \quad n \leq i \\
                n + 1           &   \mbox{if} \quad n > i
            \end{array} 
            \right. \\
    \end{array}
    \qquad
    \begin{array}{lll}
        \addOne{i}{\app{a}{b}}  & = &   \app{\addOne{i}{a}}{\addOne{i}{b}} \\
        \addOne{i}{\abs{a}}     & = &   \abs{\addOne{i+1}{a}} 
    \end{array}
\]
\end{definition}

\begin{definition}[\emph{Swap operator} -- $\swapLone{i}$] 
For every $i \in \naturalsnonzero$, $\swapLone{i}\ : \lambdaDBterm \en
\lambdaDBterm$ is given inductively by:
\[
    \begin{array}{lll}
        \swap{i}{n} & = &
            \left\{ 
            \begin{array}{ll}
                n               &   \mbox{if} \quad n < i\ \lor \  n > i + 1 \\
                i + 1           &   \mbox{if} \quad n = i \\
                i               &   \mbox{if} \quad n = i + 1
            \end{array} 
            \right.
    \end{array}
    \qquad
    \begin{array}{lll}
        \swap{i}{\app{a}{b}}    & = &   \app{\swap{i}{a}}{\swap{i}{b}} \\
        \swap{i}{\abs{a}}       & = &   \abs{\swap{i+1}{a}} 
    \end{array}
\]
\end{definition}

Finally, we present the meta-level substitution definition for $\lambdaR$, and
then the $\lambdaR$-calculus itself.

\begin{definition}[Meta-substitution for $\lambdaR$] 
For every $a, b, c \in \lambdaDBterm$, $n \in \naturalsnonzero$,
$\lambdaRsust{\funcph}{\funcph} : \lambdaDBterm \times \lambdaDBterm \en
\lambdaDBterm$ is given inductively by:
\[
    \begin{array}{lll}
        \lambdaRsust{n}{c}  & = & 
            \left\{ 
            \begin{array}{ll}
                c       &   \mbox{if} \quad n = 1 \\
                n-1     &   \mbox{if} \quad n > 1
            \end{array} 
            \right. \\
    \end{array}
    \qquad
    \begin{array}{lll}
        \lambdaRsust{(\app{a}{b})}{c}   & = &   \app{\lambdaRsust{a}{c}}{\lambdaRsust{b}{c}} \\
        \lambdaRsust{(\abs{a})}{c}      & = &   \abs{\lambdaRsust{\swap{1}{a}}{\addOne{0}{c}}}
    \end{array}
\]
\end{definition}

\begin{definition}[$\lambdaR$-calculus] 
The $\lambdaR$-calculus is the reduction system $\left( \lambdaDBterm, \betaR
\right)$, where:
\[
    \left( \forall a,b \in \lambdaDBterm \right) \left( a \betaRstep b\  \iff \left( \exists \ C\ \mbox{context};\ c, d \in \lambdaDBterm \right) 
    \left( a = \ctxt{\mathrm{C}}{(\abs{c})d} \land b = \ctxt{\mathrm{C}}{\lambdaRsust{c}{d}} \right) \right)
\]
\end{definition}

\oursubsection{$\lambdaDB$ and $\lambdaR$ are the same calculus}\label{lambda_r:sec:igualdad_lambda_db}

We want to prove that $\lambdaR$ equals $\lambdaDB$.  That is, we want to show
that $\lambdaDBsust{a}{1}{b} = \lambdaRsust{a}{b}$. In order to do this,
however, we should first prove the general case: $\lambdaDBsust{a}{n}{b} =
\lambdaRsust{a'}{b'}$, with $a'$ and $b'$ being the result of a series of
\emph{swaps} and \emph{increments} over $a$ and $b$, respectively. This comes
from observing that, while $\lambdaDB$ increments the index inside the
substitution when going into an abstraction, $\lambdaR$ performs a \emph{swap}
over the affected term, and an index increment over the term carried in the
substitution. Thus, comparing what happens after the ``crossing'' of $n-1$ abstractions
in $(\underbrace{\abs{ \cdots \abs{}}}_{n-1} \lambdaDBsust{a)}{1}{b}$ and $(\underbrace{\abs{ \cdots \abs{}}}_{n-1} \lambdaRsust{a)}{b}$, we get to:

\[\
\underbrace{\abs{ \cdots \abs{}}}_{n-1} \lambdaDBsust{a}{n}{b} \
\qquad \mbox{and} \qquad \
\underbrace{\abs{ \cdots \abs{}}}_{n-1} \lambdaRsust{\swap{1}{\cdots \swap{n-1}{a} \cdots}}{\underbrace{\addOneLone{0} \! ( \cdots \! \addOneLone{0} \! (}_{n-1} \! b ) \cdots)} \
\]

Therefore, the idea for the proof is showing that the above terms are equal for
every $n \in \naturalsnonzero$. We formalize this idea by introducing two
additional definitions: stacked swaps and stacked increments.

\begin{definition}[Stacked swap] 
For every $i \in \naturalsnonzero$, $j \in \naturals$, 
$\swapChainLone{i}{j} : \lambdaDBterm \en \lambdaDBterm$ is given inductively by:
\begin{eqnarray*}
    \swapChain{i}{j}{a}        & = &   
        \left\{ 
        \begin{array}{ll}
            a                                       &   \mbox{if $j=0$} \\
            \swapChain{i}{j-1}{\swap{i+j-1}{a}}     &   \mbox{if $j>0$}
        \end{array} 
        \right. \\
\end{eqnarray*}

The intuitive idea behind $\swapChain{i}{j}{a}$ is that of: \hspace{.4cm}
$\underbrace{\swapLone{i} \! ( \swapLone{i+1} \! ( \cdots \swapLone{i+j-1} \! (}_{j \; \mbox{\small{\emph{swaps}}}} \! a ) \cdots ))$

\end{definition}

\begin{definition}[Stacked increment] 
For every $i \in \naturals$, $\addOneChainLone{i} :
\lambdaDBterm \en \lambdaDBterm$ is given inductively by:
\begin{eqnarray*}
    \addOneChain{i}{a}        & = &   
        \left\{ 
        \begin{array}{ll}
            a                                    &   \mbox{if $i = 0$} \\
            \addOneChain{i-1}{\addOne{0}{a}}     &   \mbox{if $i > 0$}
        \end{array} 
        \right. \\
\end{eqnarray*}

The intuitive idea behind $\addOneChain{i}{a}$ is that of: \hspace{.4cm}
$\underbrace{\addOneLone{0} \! ( \cdots \! \addOneLone{0} \! ( }_{i \; \mbox{\small{increments}}} \!\!\! a ) \cdots)$

\end{definition}

Based on this last two definitions, the next theorem states the relationship
between $\lambdaR$ and $\lambdaDB$ meta-substitution operators, having as an
immediate corollary that $\lambdaR$ and $\lambdaDB$ are the same calculus.

\begin{theorem}[Correspondence between $\lambdaDB$ and $\lambdaR$ meta-substitution]\label{lambda_r:lema:correspondencia_msust}
For every $a, b \in \lambdaDBterm, \; n \in \naturalsnonzero$: 
$$\lambdaDBsust{a}{n}{b} = \; \lambdaRsust{\swapChain{1}{n-1}{a}}{\addOneChain{n-1}{b}}$$
    \begin{proof}
    See Appendix \ref{apx:lambda_r_proofs}.
    \end{proof}
\end{theorem}

\begin{cor}\label{lambda_r:cor:lambda_db_eq_lambda_r}
For every $a, b \in \lambdaDBterm : \; \lambdaDBsust{a}{1}{b} = \lambdaRsust{a}{b}$. Therefore, $\lambdaDB$ and $\lambdaR$ are the same calculus.
    \begin{proof}
    Use Theorem \ref{lambda_r:lema:correspondencia_msust} with $n = 1$, and
    conclude the equality of both calculi by definition. This result was
    checked using the \emph{Coq} theorem prover\footnote{The proof can be
    downloaded from \url{http://www.mpi-sws.org/~beta/lambdar.v}}.
    \end{proof}
\end{cor}

% Lambda_rex part
\oursection{Devising the $\lambdaRE$, $\lambdaREGC$ and $\lambdaREX$ calculi}\label{sec:lambda_rex}

In order to derive an \esabbr\ calculus from $\lambdaR$, we first need to
internalize substitutions in the language. Thus, we add the construction
$\lambdaREsust{a}{b}$ to $\lambdaDBterm$, and call the resulting set of terms
$\lambdaREterm$. The definition for the free variables of a term is extended to
consider the \esabbr\ case as follows: $\fv{\lambdaREsust{a}{b}} = \left(
\fv{a} - 1 \right) \cup \fv{b}$. Also, and as a design decision, operators
$\addOneLone{i}$ and $\swapLone{i}$ are left in the meta-level. Naturally, we
must extend their definitions to the \esabbr\ case, task that needs some lemmas
over $\lambdaR$'s meta-operators in order to ensure correctness. We use lemmas
\ref{lambda_re:lema:interaccion_msust_swaps} and
\ref{lambda_re:lema:interaccion_msust_incrementos} in Appendix
\ref{apx:extensiones_metaops} for the extension of swap and increment
meta-operators:
\[
    \addOne{i}{\lambdaREsust{a}{b}} = \, \lambdaREsust{\addOne{i+1}{a}}{\addOne{i}{b}} 
    \qquad \mbox{and} \qquad
    \swap{i}{\lambdaREsust{a}{b}} = \, \lambdaREsust{\swap{i+1}{a}}{\swap{i}{b}}
\]
Then, we just orient the equalities from the meta-substitution
definition as expected and get a calculus we call $\lambdaRE$ (that turns
out to be isomorphic to $\lambdaX$ \cite{BlRo1995,BlGe1997}, as we will later explain).

As a next step in our work, we add \emph{Garbage Collection} to $\lambdaRE$.
The goal is removing useless substitutions, \ie{when the index 1 does not
appear free in the term}. When removing a substitution, free indexes of the
term must be updated, decreasing them by 1. To accomplish this, we
introduce a new meta-operator: $\subOneLone{i}$. The operator is inspired in a
similar one from \cite{Ri1993}. We first define it for the set
$\lambdaDBterm$:

\begin{definition}[\emph{Decrement operator} -- $\subOneLone{i}$]
For every $i \in \naturalsnonzero$, $\subOneLone{i}\ : \lambdaDBterm \en \lambdaDBterm$ is given inductively by:
\[
    \begin{array}{lll}
        \subOne{i}{n}           & = &   
            \left\{ 
            \begin{array}{ll}
                n                       &   \mbox{if} \quad n < i \\
                \mbox{undefined}        &   \mbox{if} \quad n = i \\
                n - 1                   &   \mbox{if} \quad n > i
            \end{array} 
            \right. \\
    \end{array}
    \qquad
    \begin{array}{lll}
        \subOne{i}{\app{a}{b}}              & = &   \app{\subOne{i}{a}}{\subOne{i}{b}} \\
        \subOne{i}{\abs{a}}                 & = &   \abs{\subOne{i + 1}{a}}
    \end{array}
\]
    \begin{note*}
    Notice that $\subOne{i}{a}$ is well-defined iff $\,i \not\in \fv{a}$.
    \end{note*}
\end{definition}

As for the $\swapLone{i}$ and $\addOneLone{i}$ meta-operators, we need a few
lemmas to ensure a correct definition for the extension of the $\subOneLone{i}$
meta-operator to the \esabbr\ case.  Particularly, Lemma
\ref{lambda_regc:lema:distribucion_decremento} (see Appendix
\ref{apx:extensiones_metaops}) is used for this purpose. The extension
resembles those of the $\swapLone{i}$ and $\addOneLone{i}$ meta-operators:
\[
    \subOne{i}{\lambdaREsust{a}{b}} = \, \lambdaREsust{\subOne{i+1}{a}}{\subOne{i}{b}} 
\]
The \emph{Garbage Collection} rule added to $\lambdaRE$ (GC) can be seen in Figure
\ref{lambda_rex:fig:rules}, and the resulting calculus is called $\lambdaREGC$
(which, as we will see, is isomorphic to $\lambdaXGC$ \cite{BlRo1995}).

%Finally, the idea is approaching $\lambdaEX$
%\cite{Ke2008,Ke2009}. To accomplish this, we need to be able 
%to compose substitutions the $\lambdaEX$ way.
%There, composition is handled by one rule and one equation:
Finally, in order to mimic the behavior of $\lambdaEX$ \cite{Ke2009}, an
analogue method for the
composition of substitutions must be devised. In $\lambdaEX$, composition is
handled by one rule and one equation:
\vspace{-.1cm}
\[
    \begin{array}{llll}
        \lambdaXsust{\lambdaXsust{t}{x}{u}}{y}{v} &   \reducea_{\mathrm{(Comp)}}  &   \lambdaXsust{\lambdaXsust{t}{y}{v}}{x}{\lambdaXsust{u}{y}{v}}   
            &   \mbox{if} \quad y \in \fv{u} \\
        \lambdaXsust{\lambdaXsust{t}{x}{u}}{y}{v} &   =_{\mathrm{C}}  &   \lambdaXsust{\lambdaXsust{t}{y}{v}}{x}{u}                                   
            &   \mbox{if} \quad y \not\in \fv{u} \land x \not\in \fv{v}
    \end{array}
    \vspace{-.1cm}
\]
The rule (Comp) is used when substitutions are \emph{dependent}, and reasoning
modulo C-equation is needed for \emph{independent} substitutions.  Since in
$\lambdaR$-derived calculi there is no simple way of implementing an ordering
of substitutions (remember: no indexes inside closures!), and thus no trivial
path for the elimination of equation C exists, we need an analogue equation.

Let us start with the composition rule: in a term of the form
$\lambdaREsust{\lambdaREsust{a}{b}}{c}$, substitutions $[b]$ and $[c]$ are
\emph{dependent} iff $1 \in \fv{b}$. In such a term, indexes 1 and 2 in $a$ are being affected by $[b]$ and $[c]$, respectively. 
Consequently, if we were to reduce
to a term of the form $\lambdaREsust{\lambdaREsust{a'}{c'}}{b'}$, a \emph{swap}
should be performed over $a$. Moreover, as substitution $[c]$ crosses the
binder $[b]$, an index increment should also be done. Finally, since
substitutions are dependent -- that is, $[c]$ affects $b$ --, $b'$ should be
$\lambdaREsust{b}{c}$. Then, we are left with the term
$\lambdaREsust{\lambdaREsust{\swap{1}{a}}{\addOne{0}{c}}}{\lambdaREsust{b}{c}}$.

For the equation, let us suppose we negate the composition condition (\ie{$1
\not\in \fv{b}$}).  Using \emph{Garbage Collection} in the last term, we have
$\lambdaREsust{\lambdaREsust{\swap{1}{a}}{\addOne{0}{c}}}{\lambdaREsust{b}{c}}
\reducea_{\textrm{(GC)}}
\lambdaREsust{\lambdaREsust{\swap{1}{a}}{\addOne{0}{c}}}{\subOne{1}{b}}$. It is
important to notice that the condition in rule (Comp) is essential; that is: we
cannot leave (Comp) unconditional and let (GC) do its magic: we would
immediately generate infinite reductions, losing PSN. Thus, our composition
rule and equation are:
\[
    \begin{array}{llll}
        \lambdaREsust{\lambdaREsust{a}{b}}{c} &   \reducea_{\mathrm{(Comp)}}  &   \lambdaREsust{\lambdaREsust{\swap{1}{a}}{\addOne{0}{c}}}{\lambdaREsust{b}{c}}
            &   \mbox{if} \quad 1 \in \fv{b} \\
        \lambdaREsust{\lambdaREsust{a}{b}}{c} &   =_{\mathrm{D}}  &   \lambdaREsust{\lambdaREsust{\swap{1}{a}}{\addOne{0}{c}}}{\subOne{1}{b}}
            &   \mbox{if} \quad 1 \not\in \fv{b}
    \end{array}
\]

Rules for the $\lambdaREX$-calculus can be seen in Figure
\ref{lambda_rex:fig:rules}.  The relation $\REXP$ is generated by the set of
rules (App), (Lamb), (Var), (GC) and (Comp); $\lambdaREXP$ by (Beta) +
$\REXP$. D-equivalence is the least equivalence and compatible relation
generated by (EqD). Relations $\lambdaREX$ (resp. $\REX$) are obtained from
$\lambdaREXP$ (resp. $\REXP$) modulo D-equivalence (thus specifying rewriting
on D-equivalence classes). That is,
\vspace{-.2cm}
\[
    \forall \, a, a' \in \lambdaREterm : a \reducea_{(\lambda) \REX} a' 
    \; \iff \;
    \left( \exists \, b, b' \in \lambdaREterm : a \dequiv b \reducea_{(\lambda) \REXP} b' \dequiv a' \right)
    \vspace{-.1cm}
\]
We define $\lambdaREX$ as the reduction system
$(\lambdaREterm,\lambdaREX)$. We shall define $\lambdaRE$ and $\lambdaREGC$
next. Since the rule (VarR) does not belong to $\lambdaREX$, but only to
$\lambdaRE$ and $\lambdaREGC$, we present it here: 
\vspace{-.1cm}
\[ 
    \textrm{(VarR)} \qquad \lambdaREsust{(n+1)}{c} \reducea n
    \vspace{-.1cm}
\] 
The relation $\RE$ is generated by (App), (Lamb), (Var) and (VarR); $\lambdaRE$
by (Beta) + $\RE$; the relation $\REGC$ by $\RE$ + (GC); and $\lambdaREGC$ by
(Beta) + $\REGC$. Finally, the $\lambdaRE$ and $\lambdaREGC$ calculi are the
reduction systems $(\lambdaREterm, \lambdaRE)$ and $(\lambdaREterm,
\lambdaREGC)$, respectively. 

\begin{figure}[hbt]
 \center
\calculusbox{
\begin{minipage}{0.9\textwidth}
	\begin{trs}
        \trslceq{\eqdPlain}{\lambdaREsust{\lambdaREsust{a}{b}}{c}}{\lambdaREsust{\lambdaREsust{\swap{1}{a}}{\addOne{0}{c}}}{\subOne{1}{b}}}{$1 \not\in \fv{b}$}
        \trsblank

        \trslrule{Beta}{\app{(\abs{a})}{b}}{\lambdaREsust{a}{b}}
        \trslrule{App}{\lambdaREsust{(\app{a}{b})}{c}}{\app{\lambdaREsust{a}{c}}{\lambdaREsust{b}{c}}}
        \trslrule{Lamb}{\lambdaREsust{(\abs{a})}{c}}{\abs{\lambdaREsust{\swap{1}{a}}{\addOne{0}{c}}}}
        \trslrule{Var}{\lambdaREsust{1}{c}}{c}
        \trslcrule{GC}{\lambdaREsust{a}{c}}{\subOne{1}{a}}{$1 \not\in \fv{a}$}
        \trslcrule{Comp}{\lambdaREsust{\lambdaREsust{a}{b}}{c}}{\lambdaREsust{\lambdaREsust{\swap{1}{a}}{\addOne{0}{c}}}{\lambdaREsust{b}{c}}}{$1 \in \fv{b}$}
	\end{trs} 
\end{minipage}
}
\caption[$\lambdaREX$-calculus]{Equations and rules for the $\lambdaREX$-calculus}\label{lambda_rex:fig:rules}
\end{figure}

% Isomorphisms part
\oursection{The isomorphisms}\label{sec:isom}

For the isomorphism between $\lambdaEX$ and $\lambdaREX$ (and also between
$\lambdaX$ and $\lambdaRE$; and between $\lambdaXGC$ and $\lambdaREGC$), we
must first give a translation from the set $\lambdaXterm$ (\ie{the set of terms for
$\lambdaX$, $\lambdaXGC$ and $\lambdaEX$; see e.g. \cite{Ke2009} for the expected definition}) to
$\lambdaREterm$, and vice versa. It is important to notice that our translations
depend on a list of variables, which will determine the indexes of the free
variables.  All this work is inspired in a similar proof that shows the
isomorphism between the $\lambda$ and $\lambdaDB$ calculi, found in
\cite{KaRi1998}.

\begin{definition}[Translation from $\lambdaXterm$ to $\lambdaREterm$]
For every $t \in \lambdaXterm, \; n \in \naturals$, such that $\fv{t} \subseteq \left\{
x_1 , \ldots , x_n \right\}$, \\
$\tradWLone{\left[ x_1 , \ldots , x_n \right]} : \lambdaXterm \en \lambdaREterm$ is given inductively by:
\vspace{-.25cm}
\[
    \begin{array}{lll}
        \tradW{\vlste{x}{1}{n}}{x}                             & = &   \min \left\{ j : x_j = x \right\} \\
        \tradW{\vlste{x}{1}{n}}{\app{t}{u}}                    & = &   \app{\tradW{\vlste{x}{1}{n}}{t}}{\tradW{\vlste{x}{1}{n}}{u}}
    \end{array}
    \qquad
    \begin{array}{lll}
        \tradW{\vlste{x}{1}{n}}{\absv{x}{t}}                   & = &   \abs{\tradW{\left[ x, x_1, \ldots, x_n \right]}{t}} \\
        \tradW{\vlste{x}{1}{n}}{\lambdaXsust{t}{x}{u}}         & = &   \lambdaREsust{\tradW{\left[ x, x_1, \ldots, x_n \right]}{t}}{\tradW{\vlste{x}{1}{n}}{u}}
    \end{array}
\]
\end{definition}

\begin{definition}[Translation from $\lambdaREterm$ to $\lambdaXterm$]
For every $a \in \lambdaREterm, \; n \in \naturals$, such that $\fv{a} \subseteq \left\{
1 , \ldots , n \right\}$, \\
$\tradULone{\left[ x_1 , \ldots , x_n \right]} : \lambdaREterm \en
\lambdaXterm$, with $\varset{x}{1}{n}$ different variables, is given inductively
by:
\[
    \begin{array}{lll}
        \tradU{\vlste{x}{1}{n}}{j}                             & = &   x_j \\
        \tradU{\vlste{x}{1}{n}}{\app{a}{b}}                    & = &   \app{\tradU{\vlste{x}{1}{n}}{a}}{\tradU{\vlste{x}{1}{n}}{b}}
    \end{array}
    \qquad
    \begin{array}{lll}
        \tradU{\vlste{x}{1}{n}}{\abs{a}}                       & = &   \absv{x}{\tradU{\left[ x, x_1, \ldots, x_n \right]}{a}} \\
        \tradU{\vlste{x}{1}{n}}{\lambdaREsust{a}{b}}           & = &   \lambdaXsust{\tradU{\left[ x, x_1, \ldots, x_n \right]}{a}}{x}{\tradU{\vlste{x}{1}{n}}{b}}
    \end{array}
\]
with $x \not\in \varset{x}{1}{n}$ in the cases of abstraction and closure.
\end{definition}

Translations are correct w.r.t. $\alpha$-equivalence. That is,
$\alpha$-equivalent $\lambdaXterm$ terms have the same image under
$\tradWLone{\vlste{x}{1}{n}}$, and identical $\lambdaREterm$ terms have
$\alpha$-equivalent images under different choices of $x$ for
$\tradULone{\vlste{x}{1}{n}}$. Besides, adding variables at the end of
translation lists does not affect the result; thus, uniform translations
$\tradWUniLone$ and $\tradUUniLone$ can be defined straightforwardly, depending
only on a preset ordering of variables. See Appendix \ref{apx:traducciones} for
details. 

We now state the isomorphisms:

\begin{theorem}[$\lambdaEX \cong \lambdaREX$, $\lambdaX \cong \lambdaRE$ and $\lambdaXGC \cong \lambdaREGC$]\label{theo:isomorphism}
The $\lambdaEX$ (resp. $\lambdaX$, $\lambdaXGC$) and $\lambdaREX$ (resp. $\lambdaRE$, $\lambdaREGC$) calculi are isomorphic. That is,
    \begin{enumerate}[\hspace{.4cm} A.]
        \item $\comp{\tradWUniLone}{\tradUUniLone} = \textrm{Id}_{\lambdaREterm} \, \land \, \comp{\tradUUniLone}{\tradWUniLone} = \textrm{Id}_{\lambdaXterm}$
        \item $\forall t, u \in \lambdaXterm : t \reducea_{\lambdaEX (\lambdaX, \lambdaXGC)} u \implies \tradWUni{t} \reducea_{\lambdaREX (\lambdaRE, \lambdaREGC)} \tradWUni{u}$
        \item $\forall a, b \in \lambdaREterm : a \reducea_{\lambdaREX (\lambdaRE, \lambdaREGC)} b \implies \tradUUni{a} \reducea_{\lambdaEX (\lambdaX, \lambdaXGC)} \tradUUni{b}$
    \end{enumerate}

    \begin{proof}
        This is actually a three-in-one theorem. Proofs require many auxiliary
        lemmas that assert the interaction between translations and meta-operators.
        See Appendix \ref{apx:isomorphisms} for details.
    \end{proof}
\end{theorem}

\vspace{-.1cm}

Finally, in order to show meta-confluence (MC) for $\lambdaREX$, meta-variables
are added to the set of terms, and hence, functions and meta-operators are
extended accordingly. Particularly, each metavariable is decorated with a set
$\Delta$ of available free variables. This, in order to achieve an isomorphism
with $\lambdaEX$'s corresponding extension (c.f. \cite{Ke2009}). Extensions are
as follows:

\begin{enumerate}
    \item Set of terms $\lambdaREOPterm$: $a \bnfeq n \orspaced \mvar{X}{\Delta} \orspaced \app{a}{a} \orspaced \abs{a} \orspaced \lambdaREsust{a}{a} \qquad (n \in \naturalsnonzero, \, \mvar{X}{} \in \left\{ X, Y, Z, \ldots \right\}, \, \Delta \in \powerset{\naturalsnonzero})$
    \item Free variables of a metavariable: $\fv{\mvar{X}{\Delta}} = \Delta$
    \item Swap over a metavariable: $\swap{i}{\mvar{X}{\Delta}} = \mvar{X}{\Delta'}   \qquad \mbox{with } \Delta' = \setl{\Delta}{i} \cup \setg{\Delta}{i+1} \cup \left( \seteq{\Delta}{i} + 1 \right) \cup \left( \seteq{\Delta}{i+1} - 1 \right)$
    \item Increment over a metavariable: $\addOne{i}{\mvar{X}{\Delta}} = \mvar{X}{\Delta'}   \qquad \mbox{with } \Delta' = \setleq{\Delta}{i} \cup \left( \setg{\Delta}{i} + 1 \right)$
    \item Decrement over a metavariable: $\subOne{i}{\mvar{X}{\Delta}} = \left\{
        \begin{array}{ll}
            \mvar{X}{\Delta'} \qquad \mbox{with } \Delta' = \setl{\Delta}{i} \cup \left( \setg{\Delta}{i} - 1 \right)    &   \mbox{if } i \not\in \Delta \\
            \mbox{undefined}                                                                                            &   \mbox{if } i \in \Delta
        \end{array}
        \right.$
    \item Translation from $\lambdaXOPterm$ to $\lambdaREOPterm$: $ \tradW{\vlste{x}{1}{n}}{\mvar{X}{\Delta}} = \mvar{X}{\Delta'} \qquad \mbox{with } \Delta' = \{ \tradW{\vlste{x}{1}{n}}{x} : x \in \Delta \} $
    \item Translation from $\lambdaREOPterm$ to $\lambdaXOPterm$: $ \tradU{\vlste{x}{1}{n}}{\mvar{X}{\Delta}} = \mvar{X}{\Delta'} \qquad \mbox{with } \Delta' = \{ \tradU{\vlste{x}{1}{n}}{j} : j \in \Delta \} $
\end{enumerate}

\begin{theorem}\label{theo:iso_meta}
The $\lambdaREX$ and $\lambdaEX$ calculi on open terms are isomorphic.
    \begin{proof}
        This is proved as an extension of the proof for Theorem \ref{theo:isomorphism},
        considering the new case. A few simple lemmas about how meta-operators alter the
        set of free variables are needed. We refer the reader to \cite{Me2010}, chapter
        6, section 3 for details (space constraints disallow further technicality
        here).
    \end{proof}
\end{theorem}

As a direct consequence of theorems \ref{theo:isomorphism} and \ref{theo:iso_meta}, we have:

\begin{cor}[Preservation of properties]
The $\lambdaEX$ (resp. $\lambdaX$, $\lambdaXGC$) and $\lambdaREX$ (resp.
$\lambdaRE$, $\lambdaREGC$) have the same properties. In particular,
this implies $\lambdaREX$ has, among other properties, Sim, PSN and MC.
    \begin{proof}[Proof sketch for e.g. PSN in $\lambdaREX$]
        Assume PSN does not hold in $\lambdaREX$. Then, there exists $a \in
        \textrm{SN}_{\lambdaDB}$ s.t. $a \not\in \textrm{SN}_{\lambdaREX}$.
        Besides, $a \in \textrm{SN}_{\lambdaDB}$ implies $\tradUUni{a} \in
        \textrm{SN}_{\lambda}$. Therefore, by PSN of $\lambdaEX$ \cite{Ke2009}, $\tradUUni{a}
        \in \textrm{SN}_{\lambdaEX}$. Now, since $a \not\in \textrm{SN}_{\lambdaREX}$,
        there exists an infinite reduction $a \reducea_{\lambdaREX} a_1
        \reducea_{\lambdaREX} a_2 \reducea_{\lambdaREX} \cdots$. Thus, by Theorem
        \ref{theo:isomorphism}, we have $\tradUUni{a} \reducea_{\lambdaEX}
        \tradUUni{a_1} \reducea_{\lambdaEX} \tradUUni{a_2} \reducea_{\lambdaEX} \cdots$,
        contradicting the fact that $\tradUUni{a} \in \textrm{SN}_{\lambdaEX}$.
    \end{proof}
\end{cor}

% Related work
\vspace{-.3cm}
\oursection{Related work}\label{sec:related_work}

It is important to mention that, even though independently discovered, the
swapping mechanism introduced in this article was first depicted by de Bruijn
for his \esabbr\ calculus $\lambdaCXiPhi$ \cite{Br1978}, and, later, updated
w.r.t.  notation -- $\lambdaXiPhi$ -- and compared to $\lambdaUpsilon$ in
\cite{LeRo1995}. We will now briefly discuss the main differences between these
calculi and our swapping-based approach.

Firstly, neither $\lambdaCXiPhi$ nor $\lambdaXiPhi$ have composition of
substitutions nor Garbage Collection, two keys for the accomplishment of
meta-confluence. In that sense, these two calculi only resemble closely our
first $\lambdaR$-based \esabbr\ calculus: $\lambdaRE$. Thus, both $\lambdaREGC$
and $\lambdaREX$ represent a relevant innovation for swapping-based formalisms,
specially considering the fact that, as far as we know, no direct successor of
$\lambdaCXiPhi$ nor $\lambdaXiPhi$ was found to satisfy PSN and MC.

As a second fundamental difference, both $\lambdaCXiPhi$ and $\lambdaXiPhi$ are
entirely explicit formalisms. In the end, internalizing meta-operations is desirable, 
both theoretically and practically; nevertheless, the
presence of meta-operations in $\lambdaRE$, $\lambdaREGC$ and $\lambdaREX$ are
mandatory for the accomplishment of isomorphisms w.r.t. $\lambdaX$,
$\lambdaXGC$ and $\lambdaEX$, respectively.  Particularly, the isomorphism
between $\lambdaEX$ and $\lambdaREX$ represents a step forward in
the explicit substitutions area. Moreover, these
isomorphisms -- impossible in the case of $\lambdaCXiPhi$ and $\lambdaXiPhi$ --
allow simple and straightforward proofs for every single property enjoyed by
the calculi.

Last but not least, in $\lambdaCXiPhi$ as well as in $\lambdaXiPhi$, swap and
increment operations are implemented by means of a special sort of substitution
that only operates on indexes (c.f. \cite{LeRo1995}). 
%That is, swaps and increments \emph{per se} are performed only on the lower
%indexes (\ie{swaps among indexes $1$ and $2$; and increments over the index
%$1$}); and full swap and increment behavior is then accomplished by this
%special substitution type, that ``wraps'' and ``unwraps'' higher indexes until
%getting to the lower ones. 
Even though undoubtedly a very clever setting for these operations -- specially
compared to ours, much more conservative --, the fact is that we still use
meta-operations. With this in mind, it may be the case that de Bruijn's
formulation for both the swap and increment operations, if taken to the
meta-level, would lead to \emph{exactly the same} functional relations between
terms than those defined by our method.
%We have strong reasons -- not detailed here for a matter of space -- to believe
%that De Bruijn's formulation for both the swap and increment operations, if
%taken to the meta-level, would lead to the same relation between terms than
%ours. 
Consequently, this difference loses importance in the presence of
meta-operations. Nevertheless, if swap and increment meta-operations were to be
made explicit, a deep comparison between our approach and de Bruijn's should be
carried out before deciding for the use of either.
%Nevertheless, in the case of internalizing these, a deep
%comparison between both approaches when in an explicit setting should be
%performed before deciding for the use of either.

% Conclusions
\oursection{Conclusions and further work}\label{sec:conclusions}

We have presented $\lambdaREX$, an \esabbr\ calculus with \emph{de Bruijn}
indexes that is isomorphic to $\lambdaEX$, a formalism with variable names that
fulfills a whole set of interesting properties. As a consequence of the
isomorphism, $\lambdaREX$ inherits all of $\lambdaEX$'s properties. This,
together with a simple notation makes it, as far as we know, the first
calculus of its kind.  Besides, the $\lambdaRE$ and $\lambdaREGC$ calculi
(isomorphic to $\lambdaX$ and $\lambdaXGC$, respectively) were also introduced.
The development was based on a novel presentation of the classical $\lambdaDB$.
Given the homogeneity of definitions and proofs, not only for $\lambdaR$ and
$\lambdaREX$, but also for $\lambdaRE$ and $\lambdaREGC$, we think we found a
truly \emph{natural} bridge between named and indexed formalisms. We believe
this opens a new set of possibilities in the area: either by translating and
studying existing calculi with good properties; or by rethinking old calculi
from a different perspective (\ie{with $\lambdaR$'s concept in mind}). 

Work is yet to be done in order to get a more suitable theoretical tool for
implementation purposes, for unary closures and equations still make such a
task hard. In this direction, a mix of ideas from $\lambdaREX$ and calculi
with $n$-ary substitutions (\ie{$\lambdaSigma$-styled calculi}) may
lead to the solution of both issues.  Particularly, a swap-based
$\lambdaSigmaLift$ \cite{CuHaLe1996} could be an option. This comes from the
following observation: in $\lambdaSigmaLift$, the (Lamb) rule is:
\[
    \textrm{(Lamb)} \qquad  \lambdaREsust{(\abs{a})}{s} \reducea    \abs{\lambdaREsust{a}{\lift{s}}}
\]
where the intuitive semantics of $\lift{s}$ is:
$\conssigma{1}{(\compsigma{s}{\shift})}$. We observe here that \emph{this is
not nameless}! The reason is that, even though there are no explicit indexes
inside closures, this lift operation resembles closely the classic definition
of the $\lambdaDB$ calculus (particularly, leaving lower indexes untouched).
Thus, we propose replacing this rule by one of the form:
\vspace{-.1cm}
\[
    \textrm{(Lamb)} \qquad  \lambdaREsust{(\abs{a})}{s} \reducea    \abs{\lambdaREsust{\megaSwap{a}}{\lift{s}}}
    \vspace{-.1cm}
\]
with the semantics of $\lift{s}$ being $\compsigma{s}{\shift}$, and that of
$\megaSwap{a}$ being \emph{swapping} $a$'s indexes in concordance with the
substitution $s$, therefore mimicking $\lambdaR$'s behavior. This approach is
still in its early days, but we feel it is quite promising.

In a different line of work, the explicitation of meta-operators may also come
to mind: we think this is not a priority, because the main merit of
$\lambdaREX$ is evidencing the accessory nature of index updates.

From a different perspective, an attempt to use $\lambdaREX$ in proof
assistants or higher order unification \cite{DoHaKi2000} implementations may be
taken into account. In such a case, a typed version of $\lambdaREX$ should be
developed as well. Also, adding an $\eta$ rule to $\lambdaREX$ should be fairly
simple using the decrement meta-operator. Finally, studying the possible
relation between these \emph{swapping-based} formalisms and nominal logic or
nominal rewriting (see e.g. \cite{GaPi2002,FeGa2007}) could be an interesting
approach in gathering a deeper understanding of $\lambdaR$'s underlying logic.

% Acknowledgements
\vspace{.2cm}
\noindent
{\small
{\bf Acknowledgements:} Special thanks to Delia Kesner for valuable discussions and insight on the subject; as well as to the anonymous
referees for their very useful comments.
}

% Bibliography
\bibliographystyle{eptcs} % or whatever you prefer
{\tiny
\bibliography{bibliography}
}

% Appendixes
\appendix
\oursection{Proofs for the $\lambdaDB = \lambdaR$ assertion}\label{apx:lambda_r_proofs}

We first show the auxiliary lemmas that allow us to prove the main theorem
of Subsection \ref{lambda_r:sec:igualdad_lambda_db}.

\begin{lemma}\label{lambda_r:lema:addonechain_upd} 
For every $a \in \lambdaDBterm, \; n \in \naturalsnonzero, \; \addOneChain{n-1}{a} = \upd{0}{n}{a}$
    \begin{proof}
    Easy induction on $n$, using that $l \leq k < l+j \implies
    \upd{k}{i}{\upd{l}{j}{a}} = \upd{l}{j+i-1}{a}$ (c.f. \cite{KaRi1997}, lemma
    6), and the fact that $\addOne{0}{a} = \upd{0}{2}{a}$.
    \end{proof}
\end{lemma}

\begin{lemma}\label{lambda_r:lema:indices}
For every $m, i \in \naturalsnonzero$, $n \in \naturals \,$:
\begin{enumerate}
    \item \label{lambda_r:lema:indices_mayor} $m > n+i \implies \swapChain{i}{n}{m} = m$
    \item \label{lambda_r:lema:indices_menor} $i \leq m < n+i \implies \swapChain{i}{n}{m} = m+1$
    \item \label{lambda_r:lema:indices_igual} $\swapChain{i}{n}{n+i} = i$
\end{enumerate}
\begin{proof}
    Easy inductions on $n$.
\end{proof}
\end{lemma}

\begin{lemma}\label{lambda_r:lema:swapchain}
For every $a, b \in \lambdaDBterm$, $n \in \naturals$, $i \in \naturalsnonzero \,$:
\begin{enumerate}
    \item \label{lambda_r:lema:swapchain_app} $\swapChain{i}{n}{\app{a}{b}} = \; \app{\swapChain{i}{n}{a}}{\swapChain{i}{n}{b}}$
    \item \label{lambda_r:lema:swapchain_abs} $\swapChain{i}{n}{\abs{a}} = \abs{\swapChain{i+1}{n}{a}}$
    \item \label{lambda_r:lema:sust_swapchain} $\lambdaRsust{(\abs{\swapChain{2}{n}{a}})}{\addOneChain{n}{b}} = \abs{\lambdaRsust{\swapChain{1}{n+1}{a}}{\addOneChain{n+1}{b}}}$
\end{enumerate}
\begin{proof}
    Easy inductions on $n$.
\end{proof}
\end{lemma}

We now restate and prove the main theorem:

\begin{theorem*}[\textbf{\ref{lambda_r:lema:correspondencia_msust}}]
For every $a, b \in \lambdaDBterm, \; n \in \naturalsnonzero \;$, we have that
$\lambdaDBsust{a}{n}{b} = \; \lambdaRsust{\swapChain{1}{n-1}{a}}{\addOneChain{n-1}{b}}$.
    \begin{proof}
    Induction on $a$.
    \begin{itemize}
        \item $a = m \in \naturalsnonzero$. Then,
        $ 
            \lambdaDBsust{a}{n}{b} = \lambdaDBsust{m}{n}{b} = 
                \left\{
                \begin{array}{ll}
                    m-1             &   \mbox{if $m > n$} \\
                    \upd{0}{n}{b}   &   \mbox{if $m = n$} \\
                    m               &   \mbox{if $m < n$} \\
                \end{array}
                \right.
        $
        
        We consider each case separately:
        \begin{enumerate}
            \item $m > n \implies \lambdaRsust{\swapChain{1}{n-1}{m}}{\addOneChain{n-1}{b}} \stepjust{L.\ref{lambda_r:lema:indices}.\ref{lambda_r:lema:indices_mayor}}{=} \lambdaRsust{m}{\addOneChain{n-1}{b}} \stepjust{$m > n \geq 1$}{=} m-1$
            \item $m = n \implies \lambdaRsust{\swapChain{1}{n-1}{n}}{\addOneChain{n-1}{b}} \stepjust{L.\ref{lambda_r:lema:indices}.\ref{lambda_r:lema:indices_igual}}{=} \lambdaRsust{1}{\addOneChain{n-1}{b}} \stepjust{def}{=} \; \addOneChain{n-1}{b} \stepjust{L.\ref{lambda_r:lema:addonechain_upd}}{=} \upd{0}{n}{b}$
            \item $m < n \implies \lambdaRsust{\swapChain{1}{n-1}{m}}{\addOneChain{n-1}{b}} \stepjust{L.\ref{lambda_r:lema:indices}.\ref{lambda_r:lema:indices_menor}}{=} \lambdaRsust{m+1}{\addOneChain{n-1}{b}} \stepjust{$m+1 > 1$}{=} m$
        \end{enumerate}
        Then,
        \[ \lambdaDBsust{m}{n}{b} = \; \lambdaRsust{\swapChain{1}{n-1}{m}}{\addOneChain{n-1}{b}} \]

        \item $a = \app{c}{d}, \; c,d \in \lambdaDBterm$. Use inductive hypothesis and Lemma \ref{lambda_r:lema:swapchain}.\ref{lambda_r:lema:swapchain_app}.

        \item $a = \abs{c}, \; c \in \lambdaDBterm$. Then,

        \[
            \lambdaDBsust{a}{n}{b} = \lambdaDBsust{(\abs{c})}{n}{b} \stepjust{def}{=} \abs{\lambdaDBsust{c}{n+1}{b}} \stepjust{HI}{=}
            \abs{\lambdaRsust{\swapChain{1}{n}{c}}{\addOneChain{n}{b}}} \stepjust{L.\ref{lambda_r:lema:swapchain}.\ref{lambda_r:lema:sust_swapchain}}{=}
        \]
        \[
            \lambdaRsust{(\abs{\swapChain{2}{n-1}{c}})}{\addOneChain{n-1}{b}} \stepjust{L.\ref{lambda_r:lema:swapchain}.\ref{lambda_r:lema:swapchain_abs}}{=}
            \lambdaRsust{\swapChain{1}{n-1}{\abs{c}}}{\addOneChain{n-1}{b}} = \; \lambdaRsust{\swapChain{1}{n-1}{a}}{\addOneChain{n-1}{b}}
        \]

    \end{itemize}
    \end{proof}
\end{theorem*}

\oursectionshort{Extension lemmas for the $\swapLone{i}$, $\addOneLone{i}$ and $\subOneLone{i}$ meta-operators}{Extension lemmas for the meta-operators}\label{apx:extensiones_metaops}

\begin{lemma}\label{lambda_r:lema:interaccion_metaoperadores}
For every $a \in \lambdaDBterm$, $i,j \in \naturalsnonzero$, $k \in \naturals \,$: 
\begin{enumerate}
    \item $k < i \implies \swap{i+1}{\addOne{k}{a}} = \, \addOne{k}{\swap{i}{a}}$
    \item $j \geq 2 \implies \swap{i+j}{\swap{i}{a}} = \; \swap{i}{\swap{i+j}{a}}$
\end{enumerate}
\begin{proof}
    Easy induction on $a$.
\end{proof}
\end{lemma}

\begin{lemma}\label{lambda_regc:lema:subone_interacciones}
For every $a \in \lambdaDBterm$, $i \in \naturalsnonzero$, $j \in \naturals \,$:
\begin{enumerate}
    \item $i \geq j+2 \land i-1 \not\in \fv{a} \implies \subOne{i}{\addOne{j}{a}} = \, \addOne{j}{\subOne{i-1}{a}}$
    \item $j \geq 2 \land i+j \not\in \fv{a} \implies \subOne{i+j}{\swap{i}{a}} = \; \swap{i}{\subOne{i+j}{a}}$
\end{enumerate}
\begin{proof}
    Easy induction on $a$.
\end{proof}
\end{lemma}

\begin{lemma}\label{lambda_re:lema:interaccion_msust_swaps}
For every $a, b \in \lambdaDBterm$, $i \in \naturalsnonzero$, $\swap{i}{\lambdaRsust{a}{b}} = \; \lambdaRsust{\swap{i+1}{a}}{\swap{i}{b}}$
\begin{proof}
    Easy induction on $a$, using Lemma \ref{lambda_r:lema:interaccion_metaoperadores}.
\end{proof}
\end{lemma}

\begin{lemma}\label{lambda_re:lema:interaccion_msust_incrementos} 
For every $a, b \in \lambdaDBterm$, $i \in \naturals$,
$\addOne{i}{\lambdaRsust{a}{b}} = \; \lambdaRsust{\addOne{i+1}{a}}{\addOne{i}{b}}$
\begin{proof}
    Use that $\upd{k}{i}{\lambdaDBsust{a}{1}{b}} =
    \lambdaDBsust{\upd{k+1}{i}{a}}{1}{\upd{k}{i}{b}}$ (c.f. \cite{KaRi1997},
    Lemma 10 with $n = 1$), the fact that $\addOne{i}{a} = \upd{i}{2}{a}$ and
    Corollary \ref{lambda_r:cor:lambda_db_eq_lambda_r}.
\end{proof}
\end{lemma}

\begin{lemma}\label{lambda_regc:lema:distribucion_decremento} 
For every $a, b \in \lambdaDBterm$, $i \in \naturalsnonzero$,
$i+1 \not\in \fv{a} \land i \not\in \fv{b}$, we have that \\
$\subOne{i}{\lambdaRsust{a}{b}} = \; \lambdaRsust{\subOne{i+1}{a}}{\subOne{i}{b}}$.
\begin{proof}
    Easy induction on $a$, using Lemma \ref{lambda_regc:lema:subone_interacciones}.
\end{proof}
\end{lemma}

\oursection{Correction proofs for translations}\label{apx:traducciones}

We show the lemmas necessary to prove that the translations given
(\ie{$\tradWLone{\vlste{x}{1}{n}}$ and $\tradULone{\vlste{x}{1}{n}}$}) are
correct w.r.t.  $\alpha$-equivalence.

\begin{lemma}\label{lambda_re:lema:w_cambiar_variables}
For every $t \in \lambdaXterm, \; n \in \naturals$ such that $\fv{t} \subseteq \varset{x}{1}{n}$, we have that
$\forall y \not\in \left\{x_1 , \ldots , x_n \right\},$ \\ $z \in \left\{x_1 , \ldots , x_n \right\}, \;
\tradW{\vlste{x}{1}{n}}{t} = \tradW{\left[x_1 , \ldots , x_{k-1} , y , x_{k+1} , \ldots , x_n \right]}{\lambdasust{t}{z}{y}}$, 
with $k = \min \left\{ j : x_j = z \right\}$. 
    \begin{proof}
    Easy induction on $t$, but using the non-Barendregt-variable-convention
    definition for the meta-substitution operation (otherwise, we would be
    assuming that $t \alphaequiv u \implies \tradW{\vlste{x}{1}{n}}{t} =
    \tradW{\vlste{x}{1}{n}}{u}$, which is what we ultimately want to
    prove). See e.g. \cite{BlGe1997} for an expected definition.
    \end{proof}
\end{lemma}

\begin{lemma}\label{lambda_re:lema:imagen_w_cociente_alpha} 
For every $t, u \in \lambdaXterm, \; n \in \naturals$ such that $\fv{t} \subseteq \varset{x}{1}{n}$, we have that $t \alphaequiv u \implies \tradW{\vlste{x}{1}{n}}{t} =
\tradW{\vlste{x}{1}{n}}{u}$. Notice that $\tradW{\vlste{x}{1}{n}}{u}$ is well-defined, since $t \alphaequiv u \implies \fv{t} = \fv{u}$.
    \begin{proof}
    Easy induction on $t$, using Lemma
    \ref{lambda_re:lema:w_cambiar_variables}. Once again, the non-Barendregt-variable-convention 
    definition for the meta-substitution operation must be used here.
    \end{proof}
\end{lemma}

\begin{lemma}\label{lambda_re:lema:u_cambiar_variables}
For every $a \in \lambdaREterm, \; n \in \naturals, \; \varset{x}{1}{n}$ distinct variables such that $\fv{a} \subseteq
\numset{1}{n}$, we have that
$\forall y \not\in \left\{x_1 , \ldots , x_n \right\}, \; 1 \leq k \leq n :$
$\lambdasust{\tradU{\vlste{x}{1}{n}}{a}}{x_k}{y} \alphaequiv \tradU{\left[ x_1 , \ldots , x_{k-1} , y , x_{k+1} , \ldots , x_n \right]}{a}$.
    \begin{proof}
    Easy induction on $a$.
    \end{proof}
\end{lemma}

\begin{lemma}\label{lambda_re:lema:imagen_u_cociente_alpha} 
For every $a, b \in \lambdaREterm, \; n \in \naturals, \; \varset{x}{1}{n}$ distinct variables, $x, y \not\in \varset{x}{1}{n}$ such that $\fv{a} \subseteq \numset{1}{n}$, we have that:
\begin{enumerate}
    \item $\absv{x}{\tradU{\left[ x, x_1, \ldots, x_n \right]}{a}} \alphaequiv \absv{y}{\tradU{\left[ y, x_1, \ldots, x_n \right]}{a}}$
    \item $\lambdaXsust{\tradU{\left[ x, x_1, \ldots, x_n \right]}{a}}{x}{\tradU{\vlste{x}{1}{n}}{b}} \alphaequiv \lambdaXsust{\tradU{\left[ y, x_1, \ldots, x_n \right]}{a}}{y}{\tradU{\vlste{x}{1}{n}}{b}}$
\end{enumerate}
    \begin{proof}
    Direct in both cases, using the $\alpha$-equivalence definition and Lemma \ref{lambda_re:lema:u_cambiar_variables}.
    \end{proof}
\end{lemma}

Last, we show two lemmas that assert that adding variables at the end of
translation lists does not affect the result of the translation and, thus,
gives the possibility of defining uniform translations that depend only on a
preset ordering of variables.

\begin{lemma}\label{lambda_re:lema:trad_w_list_indifferent}
For every $t \in \lambdaXterm$ such that $\fv{t} \subseteq \varset{x}{1}{n}$, and for every
$\varset{y}{1}{m} \subset \mathbb{V}$, we have that $\tradW{\vlste{x}{1}{n}}{t} =
\tradW{\left[ x_1, \ldots, x_n, y_1, \ldots, y_m \right]}{t}$.
    \begin{proof}
    Easy induction on $t$.
    \end{proof}
\end{lemma}

\begin{lemma}\label{lambda_re:lema:trad_u_list_indifferent}
For every $a \in \lambdaREterm, \; \varset{x}{1}{n}$ distinct variables such that 
$\fv{a} \subseteq \numset{1}{n}$, and for every $\varset{y}{1}{m}$ distinct variables such that
$\varset{x}{1}{n} \cap \varset{y}{1}{m} = \emptyset$, we have that 
$\tradU{\vlste{x}{1}{n}}{a} \alphaequiv \tradU{\left[ x_1, \ldots, x_n, y_1, \ldots, y_m \right]}{a}$.
    \begin{proof}
    Easy induction on $a$.
    \end{proof}
\end{lemma}

\pagebreak[3]

Last, we show the definitions for uniform translations.

\begin{definition}[Uniform translation from $\lambdaXterm$ to $\lambdaREterm$]\label{lambda_re:lema:trad_w_uniform}
Given an enumeration $\left[ v_1, v_2, \ldots \right]$ of $\mathbb{V}$, 
for every $t \in \lambdaXterm, \; n \in \naturals$ such that $\fv{t} \subseteq \left\{
v_1 , \ldots , v_n \right\}$, we define $\tradWUniLone : \lambdaXterm
\en \lambdaREterm$ as: $\tradWUni{t} = \tradW{\vlste{v}{1}{n}}{t}$.
\end{definition}

\begin{definition}[Uniform translation from $\lambdaREterm$ to $\lambdaXterm$]\label{lambda_re:lema:trad_u_uniform}
Given an enumeration $\left[ v_1, v_2, \ldots \right]$ of $\mathbb{V}$,
for every $a \in \lambdaREterm, \; n \in \naturals$ such that
$\fv{a} \subseteq \numset{1}{n}$, we define $\tradUUniLone : \lambdaREterm \en \lambdaXterm$ as: $\tradUUni{a} = \tradU{\vlste{v}{1}{n}}{a}$.
\end{definition}

\oursection{Isomorphisms proofs}\label{apx:isomorphisms}

In order to prove Theorem \ref{theo:isomorphism}, we must show:

\begin{enumerate}[\hspace{.4cm} A.]
    \item \label{lambda_rex:isom:trad} $\comp{\tradWUniLone}{\tradUUniLone} = \textrm{Id}_{\lambdaREterm} \, \land \, \comp{\tradUUniLone}{\tradWUniLone} = \textrm{Id}_{\lambdaXterm}$
    \item \label{lambda_rex:isom:w_pres} $\forall t, u \in \lambdaXterm : t \reducea_{\lambdaEX (\lambdaX, \lambdaXGC)} u \implies \tradWUni{t} \reducea_{\lambdaREX (\lambdaRE, \lambdaREGC)} \tradWUni{u}$
    \item \label{lambda_rex:isom:u_pres} $\forall a, b \in \lambdaREterm : a \reducea_{\lambdaREX (\lambdaRE, \lambdaREGC)} b \implies \tradUUni{a} \reducea_{\lambdaEX (\lambdaX, \lambdaXGC)} \tradUUni{b}$
\end{enumerate}

For Part \ref{lambda_rex:isom:trad}, the following two lemmas are needed.

\begin{lemma}\label{lambda_re:lema:fv_wu} 
For every $t \in \lambdaXterm$, $a \in \lambdaREterm$, $\varset{x}{1}{n}$ variables, $\varset{y}{1}{m}$ distinct variables:
\begin{enumerate}
    \item $\fv{t} \subseteq \varset{x}{1}{n} \implies \fv{\tradW{\vlste{x}{1}{n}}{t}} \subseteq \numset{1}{n}$
    \item $\fv{a} \subseteq \numset{1}{m} \implies \fv{\tradU{\vlste{y}{1}{m}}{a}} \subseteq \varset{y}{1}{m}$
\end{enumerate}
    \begin{proof}
    \vspace{-.07cm}
    Easy inductions on $t$ and $a$, respectively.
    \end{proof}
\end{lemma}

\begin{lemma}\label{lambda_re:lema:comp} 
For every $a \in \lambdaREterm$, $t \in \lambdaXterm$: 
\begin{enumerate}
    \item $\tradWUni{\tradUUni{a}} = a$
    \item $\tradUUni{\tradWUni{t}} \alphaequiv t$
\end{enumerate}
\vspace{-.3cm}
    \begin{proof}
    Easy inductions on $a$ and $t$, respectively, using Lemma \ref{lambda_re:lema:fv_wu}.
    \end{proof}
\end{lemma}

Next, to prove Part \ref{lambda_rex:isom:w_pres} of the theorem, we need
several auxiliary lemmas, that we now state.

\begin{lemma}\label{lambda_re:lema:w_variable_no_primera_en_lista}
For every $\varset{x}{1}{n}$, $y \in \varset{x}{1}{n}$, $x \not\in \varset{x}{1}{n}$,
$\tradW{\left[ x, x_1, \ldots, x_n \right]}{y} = \tradW{\vlste{x}{1}{n}}{y} + 1$.
    \begin{proof}
    \vspace{-.07cm}
    Direct, using $\tradWUniLone$'s definition.
    \end{proof}
\end{lemma}

\begin{lemma}\label{lambda_re:lema:w_swap}
For every $t \in \lambdaXterm$, $i \in \naturalsnonzero$ such that $\fv{t} \subseteq \varset{x}{1}{n} \land i < n \land x_i \neq x_{i+1}$,\\
$\swap{i}{\tradW{\left[ x_1, \ldots, x_i, x_{i+1}, \ldots, x_n \right]}{t}} = \tradW{\left[ x_1, \ldots, x_{i+1}, x_i, \ldots, x_n \right]}{t}$.
    \begin{proof}
    \vspace{-.07cm}
    Easy induction on $t$.
    \end{proof}
\end{lemma}

\begin{lemma}\label{lambda_re:lema:w_incrementos}
For every $t \in \lambdaXterm$, $m \in \naturals$, $x \in \mathbb{V}$ such that $\fv{t} \subseteq \varset{x}{1}{n} \land m \leq n \land x \not\in \varset{x}{1}{n}$, \\
$\tradW{\left[ x_1, \ldots, x_m, x, x_{m+1}, \ldots, x_n \right]}{t} = \; \addOne{m}{\tradW{\vlste{x}{1}{n}}{t}}$.
    \begin{proof}
    \vspace{-.07cm}
    Easy induction on $t$.
    \end{proof}
\end{lemma}

\begin{lemma}\label{lambda_regc:lema:w_gc}
For every $t \in \lambdaXterm$, $m \in \naturals$, $x \in \mathbb{V}$ such that $\fv{t} \subseteq \varset{x}{1}{n} \land 1 \leq m \leq n+1$ :
    \vspace{-.07cm}
    \begin{enumerate}
        \item \label{lambda_regc:lema:w_gc:not} $x \not\in \varset{x}{1}{n} \implies m \not\in \fv{\tradW{\left[
        x_1, \ldots, x_{m-1}, x, x_m, \ldots, x_n \right]}{t}}$

        \item \label{lambda_regc:lema:w_gc:in} $x \not\in \varset{x}{1}{m-1} \land x \in \fv{t} \implies
        m \in \fv{\tradW{\left[ x_1, \ldots, x_{m-1}, x, x_m, \ldots, x_n
        \right]}{t}}$

        \item \label{lambda_regc:lema:w_gc:dec} $x \not\in \varset{x}{1}{n} \implies \tradW{\vlste{x}{1}{n}}{t} =
        \; \subOne{m}{\tradW{\left[ x_1, \ldots, x_{m-1}, x, x_m, \ldots, x_n
        \right]}{t}}$
    \end{enumerate}
    \begin{proof}
    \vspace{-.07cm}
    Easy inductions on $t$.
    \end{proof}
\end{lemma}

Given the auxiliary lemmas, we proceed to prove Part
\ref{lambda_rex:isom:w_pres} of the isomorphism theorem. Item
\ref{lambda_re:lema:mantenimiento_reduccion_w_red}
of the next lemma is enough to prove the reduction preservation under
translation $\tradWUniLone$ for the $\lambdaRE$ and $\lambdaREGC$ calculi. For
$\lambdaREX$, Item \ref{lambda_re:lema:mantenimiento_reduccion_w_eqr}
-- showing the preservation of the equivalence relations under
translation $\tradWUniLone$ -- is also needed. Then, preservation for
$\lambdaREX$ follows immediately from the definition of reduction modulo an
equivalence relation.

\begin{lemma}\label{lambda_re:lema:mantenimiento_reduccion_w}
For every $t, u \in \lambdaXterm :$
    \begin{enumerate}
        \item \label{lambda_re:lema:mantenimiento_reduccion_w_red} $t
        \reducea_{\BX (\lambdaX, \lambdaXGC)} u \implies \tradWUni{t}
        \reducea_{\lambdaREXP (\lambdaRE, \lambdaREGC)} \tradWUni{u}$

        \item \label{lambda_re:lema:mantenimiento_reduccion_w_eqr} $t
        \cequiv u \implies \tradWUni{t} \dequiv \tradWUni{u}$
    \end{enumerate}
    \begin{proof}
    \textbf{Part \ref{lambda_re:lema:mantenimiento_reduccion_w_red}.}
    Induction on $t$. The only interesting cases are those of the explicit
    substitution when the reduction takes place at the root. The rest of the
    cases are either trivial or easily shown by using the inductive hypothesis.
    %We will show here the explicit substitution cases in which reduction is
    %done by using the (Lamb), (GC) or (Comp) rules. Thus, $t$ is of the form
    %$\lambdaXsust{t_1}{x}{t_2}$.
    We will show the explicit substitution case in which reduction is done
    by using the (Comp) rule. The other two relevant cases, (Lamb) and (GC),
    omitted here for a matter of space, are proved in a similar fashion. Since
    we are working in the explicit substitution case, $t$ is of the form
    $\lambdaXsust{t_1}{x}{t_2}$. Now, as the (Comp) rule is used, we have that:
    $$\lambdaXsust{t_1}{x}{t_2} = \lambdaXsust{\lambdaXsust{t_3}{y}{t_4}}{x}{t_2} \stepjust{(Comp)}{\BXstep} \;
    \lambdaXsust{\lambdaXsust{t_3}{x}{t_2}}{y}{\lambdaXsust{t_4}{x}{t_2}} = u$$
    with $x \in \fv{t_4}$. By the variable convention, we assume $x
    \neq y \land y \not\in \varset{x}{1}{n}$. Thus,
    \[
        \tradW{\vlste{x}{1}{n}}{\lambdaXsust{\lambdaXsust{t_3}{y}{t_4}}{x}{t_2}} \stepjust{def}{=}
        \lambdaREsust{\tradW{\left[ x, x_1, \ldots, x_n \right]}{\lambdaXsust{t_3}{y}{t_4}}}{\tradW{\vlste{x}{1}{n}}{t_2}}
        \stepjust{def}{=}
    \]
    \[
        \lambdaREsust{\lambdaREsust{\tradW{\left[ y, x, x_1, \ldots, x_n \right]}{t_3}}{\tradW{\left[ x, x_1, \ldots, x_n \right]}{t_4}}}{\tradW{\vlste{x}{1}{n}}{t_2}}
        \stepjust{L.\ref{lambda_regc:lema:w_gc}.\ref{lambda_regc:lema:w_gc:in}, $x \in \fv{t_4}$, (Comp)}{\lambdaREXPstep}
    \]
    \[
        \lambdaREsust{\lambdaREsust{\swap{1}{\tradW{\left[ y, x, x_1, \ldots, x_n \right]}{t_3}}}{\addOne{0}{\tradW{\vlste{x}{1}{n}}{t_2}}}}{\lambdaREsust{\tradW{\left[ x, x_1, \ldots, x_n \right]}{t_4}}{\tradW{\vlste{x}{1}{n}}{t_2}}}
        \stepjust{L.\ref{lambda_re:lema:w_swap}, $x \neq y$}{=}
    \]
    \[
        \lambdaREsust{\lambdaREsust{\tradW{\left[ x, y, x_1, \ldots, x_n \right]}{t_3}}{\addOne{0}{\tradW{\vlste{x}{1}{n}}{t_2}}}}{\lambdaREsust{\tradW{\left[ x, x_1, \ldots, x_n \right]}{t_4}}{\tradW{\vlste{x}{1}{n}}{t_2}}}
        \stepjust{L.\ref{lambda_re:lema:w_incrementos}, $y \not\in \fv{t_2}$}{=}
    \]
    \[
        \lambdaREsust{\lambdaREsust{\tradW{\left[ x, y, x_1, \ldots, x_n \right]}{t_3}}{\tradW{\left[ y, x_1, \ldots, x_n \right]}{t_2}}}{\lambdaREsust{\tradW{\left[ x, x_1, \ldots, x_n \right]}{t_4}}{\tradW{\vlste{x}{1}{n}}{t_2}}}
        \stepjust{def}{=}
    \]
    \[
        \lambdaREsust{\lambdaREsust{\tradW{\left[ x, y, x_1, \ldots, x_n \right]}{t_3}}{\tradW{\left[ y, x_1, \ldots, x_n \right]}{t_2}}}{\tradW{\vlste{x}{1}{n}}{\lambdaXsust{t_4}{x}{t_2}}}
        \stepjust{def}{=}
    \]
    \[
        \lambdaREsust{\tradW{\left[ y, x_1, \ldots, x_n \right]}{\lambdaXsust{t_3}{x}{t_2}}}{\tradW{\vlste{x}{1}{n}}{\lambdaXsust{t_4}{x}{t_2}}}
        \stepjust{def}{=}
        \tradW{\vlste{x}{1}{n}}{\lambdaXsust{\lambdaXsust{t_3}{x}{t_2}}{y}{\lambdaXsust{t_4}{x}{t_2}}}
    \]

    \textbf{Part \ref{lambda_re:lema:mantenimiento_reduccion_w_eqr}.} 
    Induction on the inference of $t \cequiv u$. The only interesting case is
    when the actual equation is used.  Then, $t =
    \lambdaXsust{\lambdaXsust{t_1}{y}{t_2}}{x}{t_3} \cequiv
    \lambdaXsust{\lambdaXsust{t_1}{x}{t_3}}{y}{t_2} = u$, with $x \neq y \land
    x \not\in \fv{t_2} \land y \not\in \fv{t_3}$.  By the variable convention,
    assume that $\left\{ x, y \right\} \cap \varset{x}{1}{n} = \emptyset$.
    %So,
    Proceed in a similar way than that of the proof of Part
    \ref{lambda_re:lema:mantenimiento_reduccion_w_red} in the (Comp) case.
    \end{proof}
\end{lemma}

Finally, to prove Part \ref{lambda_rex:isom:u_pres} of the isomorphism theorem,
we also need several auxiliary lemmas analogue to those used for Part
\ref{lambda_rex:isom:w_pres}. We will now state them.

\begin{lemma}\label{lambda_re:lema:u_swap}
For every $a \in \lambdaREterm$, $i \in \naturalsnonzero$, $\varset{x}{1}{n}$ distinct variables such that $\fv{a} \subseteq \numset{1}{n} \\ \land i < n$, we have that 
$\tradU{\left[ x_1, \ldots, x_i, x_{i+1}, \ldots, x_n \right]}{a} \alphaequiv \tradU{\left[ x_1, \ldots, x_{i+1}, x_i, \ldots, x_n \right]}{\swap{i}{a}}$.
    \begin{proof}
    Easy induction on $a$.
    \end{proof}
\end{lemma}

\begin{lemma}\label{lambda_re:lema:u_incrementos}
For every $a \in \lambdaREterm$, $m \in \naturals$, $\varset{x}{1}{n}$ distinct variables, $x \not\in \varset{x}{1}{n}$ such that \\
$\fv{a} \subseteq \numset{1}{n} \land m \leq n$, we have that
$\tradU{\left[ x_1, \ldots, x_m, x, x_{m+1}, \ldots, x_n \right]}{\addOne{m}{a}} \alphaequiv \tradU{\vlste{x}{1}{n}}{a}$.
    \begin{proof}
    Easy induction on $a$.
    \end{proof}
\end{lemma}

\begin{lemma}\label{lambda_regc:lema:u_gc}
For every $a \in \lambdaREterm$, $m \in \naturals$, $x \in \mathbb{V}$,
$\varset{x}{1}{n}$ distinct variables such that $\fv{a} \subseteq \numset{1}{n} \land 1
\leq m \leq n+1 \land x \not\in \varset{x}{1}{n} \,$:
    \begin{enumerate}
        \item \label{lambda_regc:lema:u_gc:not} $m \not\in \fv{a} \implies x \not\in \fv{\tradU{\left[ x_1,
        \ldots, x_{m-1}, x, x_m, \ldots, x_n \right]}{a}}$

        \item \label{lambda_regc:lema:u_gc:in} $m \in \fv{a} \implies x \in \fv{\tradU{\left[ x_1,
        \ldots, x_{m-1}, x, x_m, \ldots, x_n \right]}{a}}$

        \item \label{lambda_regc:lema:u_gc:dec} $m \not\in \fv{a} \implies \tradU{\left[ x_1, \ldots, x_{m-1}, x,
        x_m, \ldots, x_n \right]}{a} \alphaequiv
        \tradU{\vlste{x}{1}{n}}{\subOne{m}{a}}$
    \end{enumerate}
    \begin{proof}
    Easy inductions on $a$.
    \end{proof}
\end{lemma}

Given, once again, the auxiliary lemmas, we will now state Part
\ref{lambda_rex:isom:u_pres} of the isomorphism theorem. As for Part
\ref{lambda_rex:isom:w_pres}, Item
\ref{lambda_re:lema:mantenimiento_reduccion_u_red} of Lemma
\ref{lambda_re:lema:mantenimiento_reduccion_u} will be enough to prove
preservation for the $\lambdaX$ and $\lambdaXGC$ calculi, whereas Item
\ref{lambda_re:lema:mantenimiento_reduccion_u_eqr} will also be needed for the
case of $\lambdaEX$, concluding preservation by definition of reduction modulo
an equation.

\begin{lemma}\label{lambda_re:lema:mantenimiento_reduccion_u}
For every $a, b \in \lambdaREterm :$
    \begin{enumerate}
        \item \label{lambda_re:lema:mantenimiento_reduccion_u_red} $a \reducea_{\lambdaREXP (\lambdaRE, \lambdaREGC)} b \implies \tradUUni{a} \reducea_{\BX (\lambdaX, \lambdaXGC)} \tradUUni{b}$
        \item \label{lambda_re:lema:mantenimiento_reduccion_u_eqr} $a \dequiv b \implies \tradUUni{a} \cequiv \tradUUni{b}$
    \end{enumerate}
    
    \begin{proof}
    For part \ref{lambda_re:lema:mantenimiento_reduccion_u_red}, perform induction on $a$ analogue to that
    of lemma \ref{lambda_re:lema:mantenimiento_reduccion_w}.\ref{lambda_re:lema:mantenimiento_reduccion_w_red}. 
    For part \ref{lambda_re:lema:mantenimiento_reduccion_u_eqr}, perform induction on the inference of $a \dequiv b$, analogue to
    that of lemma \ref{lambda_re:lema:mantenimiento_reduccion_w}.\ref{lambda_re:lema:mantenimiento_reduccion_w_eqr}. In both cases,
    use auxiliary lemmas \ref{lambda_re:lema:u_swap}, \ref{lambda_re:lema:u_incrementos} and \ref{lambda_regc:lema:u_gc}.
    \end{proof}

\end{lemma}

\end{document}